\documentclass[amsmath,amssymb,amsfonts,aps,pre,preprint,superscriptaddress,bibnotes,showpacs,showkeys,longbibliography]{revtex4-1}

\usepackage{graphicx}
\usepackage{subfigure}
\usepackage{amsmath}
\usepackage{amsfonts}
\usepackage{amssymb}
\usepackage{natbib}
\usepackage{relsize}
\usepackage{sidecap}
\usepackage{braket}
\usepackage{footmisc}
\usepackage{footnote}
\usepackage{color}
\usepackage{hyperref}
\begin{document}

\title{Static interfacial properties of Bose-Einstein condensate mixtures}

\author{Joseph O. Indekeu}
\affiliation{Institute for Theoretical Physics, KU Leuven, Celestijnenlaan 200D,  B-3001 Leuven, Belgium}

\author{Chang-You Lin}
\affiliation{Institute for Theoretical Physics, KU Leuven, Celestijnenlaan 200D,  B-3001 Leuven, Belgium}

\author{Nguyen Van Thu}
\affiliation{Department of Physics, Hanoi Pedagogical University No. 2, Hanoi, Vietnam}

\author{Bert Van Schaeybroeck}
\affiliation{Royal Meteorological Institute, Ringlaan 3, 1180 Brussels, Belgium.}

\author{Tran Huu Phat}
\affiliation{Vietnam Atomic Energy Commission, 59 Ly Thuong Kiet, Hanoi, Vietnam}

\begin{abstract}
Interfacial profiles and interfacial tensions of phase-separated binary mixtures of Bose-Einstein condensates are studied theoretically. The two condensates are characterized by their respective healing lengths $\xi_1$ and $\xi_2$ and by the inter-species repulsive interaction $K$. An exact solution to the Gross-Pitaevskii (GP) equations is obtained for the special case $\xi_2/\xi_1$ = 1/2 and $K = 3/2$. Furthermore, applying a double-parabola approximation (DPA) to the energy density featured in  GP theory allows us to define a DPA model, which is much simpler to handle than GP theory but nevertheless still captures the main physics. In particular, a compact analytic expression for the interfacial tension is derived that is useful for all $\xi_1, \xi_2$ and $K$.  An application to wetting phenomena is presented for condensates adsorbed at an optical wall. The wetting phase boundary obtained within the DPA model nearly coincides with the exact one in GP theory.
\end{abstract}

\maketitle

\section{Introduction}
Phase separation in gas mixtures goes against our intuitive notion of the entropy of mixing. However, at ultralow temperature where quantum physics rules, this possibility has been realized experimentally, in particular in binary mixtures of Bose-Einstein condensates (BECs). See, e.g., \cite{UBFG} for an overview. Let us recall briefly the state-of-the-art research in this field, paying special attention to experiments.

At the beginning of this century weakly phase-segregated binary Bose-Einstein systems have been observed experimentally~\cite{modugno,miesner,myatt,stamper2,hall,matthews}. More recently, strong phase separation  has been realized  by various research groups~\cite{mccarron,tojo,altin,papp2,xiong}. Moreover, also in an ultracold  dual component gas of thermal atoms spatial separation of components has been achieved~\cite{baumer}, while many more degenerate Bose mixtures were produced in which phase separation is possible~\cite{stellmer,pilch,thalhammer}. The theoretical and experimental physics of multi-component condensates is well explained in Refs.~\cite{stamper3,malomed2}. The statics and dynamics of phase separated BECs have been extensively addressed in Refs.~\cite{ho,ejnisman,pu,alexandrov,timmermans,ao1998binary,svidzinsky,svidzinsky2,navarro,wen,ronen,pattinson,gautam}. The interfacial phenomomenology in Bose mixtures has been explored in Refs.~\cite{mazets,bezett,sasaki,ticknor,takeuchi,goldman,kadokura,pepe} and the phase diagram at finite temperature was investigated in Refs.~\cite{phat,BVS2,shi}.

Our focus in this paper is on the calculation of static interfacial properties of BEC binary mixtures within Gross-Pitaevskii (GP) theory \cite{GP}. Research on this problem, in particular on the interfacial tension of phase-separated mixtures of BECs  \cite{ao1998binary}, has led to interesting results \cite{timmermans,BertVan}. In particular, in \cite{BertVan}  accurate analytical approximations (e.g., series expansions) have been provided, covering certain ranges of condensate healing lengths  and  interparticle repulsive interaction strengths. While these results are useful, there is still a need for i) more exact solutions within GP theory, and ii) a simpler model which can provide a compact and insightful expression for the interfacial tension that can be used over the entire parameter range.  Our aim is to contribute advances to meet both of these needs.

This paper is organized as follows. Section II deals with the frame-work of GP theory. Section III is concerned with exact solutions to the GP equations describing interfacial profiles and announces our exact solution for a special choice of physical parameters. Section IV defines our simpler model through the so-called double-parabola approximation (DPA) to the GP Lagrangian and presents the solutions for the interfacial profiles within this model. Section V treats the application of the DPA to the interfacial tension and presents our compact analytical expression for this quantity. Section VI is concerned with the application of the DPA to the wetting phase transition that can occur when the condensates are adsorbed at an optical wall. This section also clarifies how the exact solution announced in section II could be found by virtue of insights gained through the DPA. Finally, our conclusions and outlook are given in Section VII.

\section{Gross-Pitaevskii theoretical frame-work}

\subsection{Gross-Pitaevskii Lagrangian}

We start from the Lagrangian $L$ and action $S$ of a two-component BEC
\begin{equation}\label{eq:action}
S \left[\Psi_1,\Psi_2 \right]= \int \mathrm{d}t L =\int \mathrm{d}t \mathrm{d} \mathbf{r} \mathcal{L}
\end{equation}
with the mean-field Gross-Pitaevskii Lagrangian density \cite{pethick2002bose,PitaString}
\begin{equation}\label{eq:lagrangiandensity}
\mathcal{L} (\Psi_1,\Psi_2 )=  \sum_{j=1}^{2}  {\color{black}\frac{i\hbar}{2} \left(\Psi_j^{\ast} \partial_t \Psi_j -\Psi_j \partial_t \Psi_j^{\ast} \right)} - \mathcal{E}(\Psi_1,\Psi_2),
\end{equation}
in which the Hamiltonian density is
\begin{equation}\label{eq:hamiltoniandensity}
\mathcal{E} (\Psi_1,\Psi_2)=  \sum_{j=1}^{2} \left[ {\color{black} \frac{\hbar^2}{2 m_j}  \left|\nabla \Psi_j \right|^2 } + \frac{g_{jj}}{2}  \lvert\Psi_j\rvert^4  \right] + g_{12} \lvert\Psi_1\rvert^2 \lvert\Psi_2\rvert^2,
\end{equation}
where, for species $j$, $\Psi_j=\Psi_{j}(\mathbf{r},t)$ is the wave function of the condensate playing the role of order parameter; $m_j$ is the atomic mass, $g_{jj}=4 \pi \hbar^2 a_{jj}/m_j >0$ is the strength of the repulsive intra-species interaction, $g_{12}=2 \pi \hbar^2 a_{12}(1/m_1 + 1/m_2)>0$ is the strength of the repulsive inter-species interaction and $a_{jj'}$ is the $s$-wave scattering length, relevant at low energies. 

By introducing the following dimensionless quantities, $\mathbf{s}_j=\mathbf{r}/\xi_j$, with $\xi_j = \hbar/\sqrt{2m_jn_{j0}g_{jj}}$ the healing length and $n_{j0}$ the number density of condensate $j$ in bulk, $\tau_j=t/t_{j}$, $\psi_{j}=\Psi_{j}/\sqrt{n_{j0}}$, and $K=g_{12}/\sqrt{g_{11} g_{22}}$, where $t_{j}=\hbar/\mu_j$, and $\mu_j=g_{jj} n_{j0}$ the chemical potential of condensate $j$, we scale the Lagrangian density in \eqref{eq:lagrangiandensity} and Hamiltonian density in \eqref{eq:hamiltoniandensity} to
\begin{equation}\label{eq:dimensionlesslagrangiandensity}
\tilde{\mathcal{L}} (\psi_1,\psi_2) = \frac{\mathcal{L}}{2 P_0}=  \sum_{j=1}^{2} {\color{black}\frac{i}{2} \left(\psi_j^{\ast} \partial_{\tau_j} \psi_j  - \psi_j \partial_{\tau_j} \psi_j^{\ast} \right)} - \tilde{\mathcal{E}} (\psi_1,\psi_2),
\end{equation}
with
\begin{equation}
\tilde{\mathcal{E}} (\psi_1,\psi_2)= \frac{\mathcal{E}}{2 P_0} = \sum_{j=1}^{2} \left[ {\color{black} \left| \nabla_{\mathbf{s}_j} \psi_j \right|^2 }+ \frac{\lvert\psi_j\rvert^4}{2} \right] + K \lvert\psi_1\rvert^2 \lvert\psi_2\rvert^2,
\end{equation}
where the pressure $P_0$ is given by $\mu_j^2/2 g_{jj}$, which takes one and the same value in both condensates at two-phase coexistence.
Next we make a transformation of the dimensionless Lagrangian density by writing 
\begin{equation}
\psi_j (\mathbf{s}_j,\tau_j) \equiv \phi_j (\mathbf{s}_j,\tau_j)e^{- i\tau_j }.
\end{equation}
We then have a Lagrangian density in terms of the new order parameters $\phi_j$, 
\begin{equation}\label{eq:dimensionlesslagrangiandensityinphi}
\hat{\mathcal{L}} \left(\phi_1,\phi_2 \right) \equiv \tilde{\mathcal{L}} \left(\phi_1 e^{- i\tau_1}, \phi_2 e^{- i\tau_2} \right) = \sum_{j=1}^{2}  \left[  {\color{black} \frac{i}{2} \left( \phi_j^{\ast}  \partial_{\tau_j} \phi_j - \phi_j \partial_{\tau_j} \phi_j^{\ast} \right) - \left|\nabla_{\mathbf{s}_j} \phi_j \right|^2 } \right] - \hat{\mathcal{V}} (\phi_1,\phi_2),
\end{equation}
in which the potential $\tilde{\mathcal{V}}$ takes the form
\begin{equation}\label{eq:potential}
\tilde{\mathcal{V}}(\phi_1,\phi_2)= \sum_{j=1}^{2}  \left[-\lvert\phi_j\rvert^2 + \frac{\lvert\phi_j\rvert^4}{2} \right] + K \lvert\phi_1\rvert^2 \lvert\phi_2\rvert^2.
\end{equation}
We recall that when $K>1$, the two components are immiscible and a two-phase segregated BEC is formed \cite{ao1998binary}.

\subsection{GP Equations}

By considering a variation $\phi_j^{\ast} \rightarrow \phi_j^{\ast} + \delta \phi_j^{\ast}$ and requiring $\delta S/\delta  \phi_j^{\ast} =0$, we obtain the Euler-Lagrange equations 
\begin{equation}\label{eq:ELeq}
\frac{\partial\hat{\mathcal{L}}}{\partial \phi_j^{{\color{black}\ast}}} = \partial_{\tau_j} \frac{\partial\hat{\mathcal{L}}}{\partial (\partial_{\tau_j} \phi_j^{{\color{black}\ast}})};\, j=1,2,
\end{equation}
yielding the time-dependent GP equations
\begin{equation}
i \partial_{\tau_j} \phi_j = - \nabla^2_{\mathbf{s}_j} \phi_j +  \frac{\partial \tilde{\mathcal{V}}}{\partial \phi_j^{\ast} };\, j=1,2
\end{equation}
or
\begin{equation}\label{eq:GPE}
i \partial_{\tau_j} \phi_j  = \left[- \nabla_{\mathbf{s}_j}^2 -1 + \lvert\phi_j\rvert^2 + K \lvert\phi_{j'}\rvert^2 \right] \phi_j;  \;\; j=1,2 \; (j \ne j').
\end{equation}
Note that these reduce to the time-independent GP equations (TIGPE) when the order parameter $\phi_j (\mathbf{s}_j,\tau_j) = \phi_{j0}(\mathbf{s}_j)$ is time-independent (so that $\psi_j (\mathbf{s}_j,\tau_j) =  \phi_{j0}(\mathbf{s}_j)e^{- i\tau_j }$ is stationary), namely
\begin{equation}\label{eq:TIGPEgeneralV}
 \nabla^2_{\mathbf{s}_j} \phi_{j0} =   \frac{\partial \tilde{\mathcal{V}}}{\partial \phi_{j0}^{\ast} };\, j=1,2,
\end{equation}
which leads to
\begin{equation}\label{eq:TIGPE}
\left[- \nabla_{\mathbf{s}_j}^2 -1 + \lvert\phi_{j0}\rvert^2 + K \lvert\phi_{j'0}\rvert^2 \right] \phi_{j0} = 0;  \;\; j=1,2 \; (j \ne j').
\end{equation}

\subsection{Boundary conditions for interface profiles}

For describing a static planar interface at $z=0$, separating bulk condensate 1 residing in the half-space $z \ge 0$ and bulk condensate 2 residing in the half-space $z \le 0$, we limit our attention to order parameters that are translationally invariant in the $x$ and $y$ directions. To keep the notation simple, we will not change the name of the function:  $\phi_{j0}(\mathbf{s}_j) \rightarrow \phi_{j0}( \rho_j)$, where $\rho_j \equiv z/\xi_j$. For describing an interface the TIGPE must be solved with the boundary conditions
\begin{equation}\label{eq:planarbc}
\begin{split}
\phi_{10}(\rho_1\rightarrow \infty) &=\phi_{20}(\rho_2\rightarrow-\infty)=1 \\
\phi_{20}(\rho_2\rightarrow\infty) & =\phi_{10}(\rho_1\rightarrow-\infty) =0
\end{split}
\end{equation}

\section{Exact solutions for interfacial profiles}

\subsection{The strong segregation limit $K \rightarrow \infty$}

In the interesting limit of strong inter-species repulsion $K \rightarrow \infty$, the segregation (i.e., mutual exclusion in space) of the phases becomes complete. Numerical solution of the TIGPE indicates that the overlap of two order parameters becomes zero and that the interaction term $K \lvert\phi_1\rvert^2 \lvert\phi_2\rvert^2$ in the potential \eqref{eq:potential} becomes negligible. Consequently, the GP equations decouple in this limit (in spite of the divergence of $K$). In this limit the simple exact solution to the GP equations for the interface consists of a pair of adjacent ``tanh" profiles
\begin{equation}\label{eq:stationarysolutionsGPE}
\phi_{j0} (\rho_j) = \tanh\left[ (-1)^{j+1}\frac{\rho_j}{\sqrt{2}}\right]; \; j=1,2.
\end{equation}
The interface position is conveniently marked by the common point of vanishing of the two order parameters. Note that the interface consists simply of two adjacent ``hard wall" profiles \cite{Fetter}. It is noteworthy that a (single) condensate wave function at a hard wall is mathematically similar to the Ginzburg-Landau (GL) order parameter for a strongly type-I superconductor at a normal/superconducting interface in the limit $\kappa \rightarrow 0$, with $\kappa$ the GL parameter \cite{GL}. 

At this point we make two remarks. Firstly, the fact that the interaction term $K \lvert\phi_1\rvert^2 \lvert\phi_2\rvert^2$ in the potential \eqref{eq:potential} becomes negligible, can be understood analytically using the DPA, which will be introduced in the next section. Secondly, the appearance of the ``tanh" function is not unique to the strong segregation limit. In the next subsections we will see two more examples of the occurrence of this function in exact solutions for finite $K$.

\subsection{The exact solution of Malomed {\em et al.}}
Interestingly, for $K=3$ and for the symmetric choice $\xi_1=\xi_2$, an exact solution to coupled differential equations mathematically identical to the GP equations was provided by Malomed {\em et al.} \cite{Malomed} in the physical context of domain boundaries in convection patterns. The solutions are surprisingly simple and again involve the ``tanh" function,
\begin{equation}\label{Malomed}
\phi_{10}= \frac{1}{2} \left[ 1+ \tanh \left( \frac{z}{\sqrt{2} \,\xi}\right) \right]  ; \;\;\phi_{20}= \frac{1}{2} \left[ 1- \tanh \left( \frac{z}{\sqrt{2} \,\xi}\right ) \right].
\end{equation} 

We will come back to this special case when we discuss the interfacial tension in section V. There it will become clear that the DPA can explain why we might have anticipated that the exact solution is simple for this special case $K=3$ and  $\xi_1=\xi_2$.

\subsection{An exact solution for an asymmetric case ($\xi_1\neq \xi_2$)}

We have uncovered another exact solution. It applies to the case  $\xi_2/\xi_1 = 1/2$ and $K=3/2$ (the roles of $\xi_1$ and $\xi_2$ can be interchanged). The  solutions to the GP equations are now the following profiles, with $\xi_1=\xi$, and $\xi_2 =\xi/2$, once more involving the ``tanh" function,
\begin{equation}\label{serendipity}
\phi_{10}= \sqrt{\frac{1}{2} \left[ 1+ \tanh \left( \frac{z}{\sqrt{2}\, \xi}  \right) \right] } ; \; \;\phi_{20}= \frac{1}{2} \left[ 1- \tanh \left( \frac{z}{\sqrt{2} \,\xi} \right) \right].
\end{equation}
The heuristic procedure through which this serendipitous solution could be  found has been strongly guided by insights provided by the DPA. We will discuss this in detail in section VI when we treat the wetting problem within DPA. In this and the preceding two subsections we have disclosed that the DPA is useful for guessing exact results. We now turn to the precise definition of the DPA in the context of GP theory and to the study of the remarkable properties of the simple model defined through the DPA. 

\section{Double-Parabola Approximation (DPA) for interfacial profiles}

The idea of approximating a double-well potential by a piece-wise parabolic function is old and generally dictated by the desire to work with piece-wise harmonic potentials that allow one to solve the equations of motion exactly using simple functions, the behavior of which is easy to interpret physically. In the context of surface and interfacial phenomena one of the first to implement this approximation for the bulk free-energy density was Hauge \cite{Hauge}. We follow this line of thought and expand the quartic potential $\hat{\mathcal{V}}$ in \eqref{eq:potential} about its (local) minima, which correspond to the bulk values for the order parameters. For obtaining the interface profile in the half-space $z <0$ we make use of the expansion about bulk condensate 2 and for the half-space $z >0$ we expand about condensate 1, in the following manner,

\begin{equation}\label{eq:DPA}
{\color{black}\left|\phi_j\right|}=1+\epsilon_j; \; {\color{black}\left|\phi_{j'}\right|} = \delta_{j'},
\; \mbox{with} \;
\begin{cases}
z \ge 0, & (j,j')=(1,2) \\
z \le 0, & (j,j')=(2,1)
\end{cases}
\end{equation}
in which the real numbers  $\epsilon_j $ and $\delta_j $ are treated as small perturbations. We expand the potential up to second order in $\epsilon_j$ and $\delta_j$ and so arrive at two ``quadratic"  potentials, each of which is to be used in the appropriate half-space,

\begin{equation}\label{eq:DPApotential}
\hat{\mathcal{V}}_{\mathrm{DPA}}(\phi_1,\phi_2) =  2\left(\left|\phi_j\right|  -1\right)^2 + (K-1) \lvert \phi_{j'} \rvert^2 -\frac{1}{2},
\; \mbox{with} \;
\begin{cases}
z \ge 0, & (j,j')=(1,2) \\
z \le 0, & (j,j')=(2,1)
\end{cases}
\end{equation}
This defines the DPA for the potential energy density and can be interpreted as the potential for a model that is related to, but different from, the original GP theory and which we will call the DPA model. The following technical remark is in order: In view of the structure of the TIGPE we anticipate that, at the interface position $z=0$, it will (in general) be possible to preserve continuity of the order parameter functions {\em and} their first derivatives. Indeed, in  view of \eqref{eq:TIGPEgeneralV}, continuity of the potential $\hat{\mathcal{V}}$ but discontinuity of one of its (partial) derivatives with respect to the order parameter(s), at $z=0$, will induce a {\em discontinuity in the second derivative} of the interface profile functions $\phi_{j0} (\rho_j)$; $j=1,2$. This is a mild singularity, often imperceptible to the eye in an interface plot. Experience has taught us that this singularity has little or no effect on the qualitative features of the phenomena under study, provided one stays away from bulk criticality or similar conditions, at which the (local) minima of the potential may merge or undergo some other drastic change.

\subsection{DPA for the GP equations}

The double-parabola-approximated GP equations are obtained by replacing the potential $\hat{\mathcal{V}}$ in the Lagrangian density \eqref{eq:dimensionlesslagrangiandensityinphi} by the $\hat{\mathcal{V}}_{\mathrm{DPA}}$ in \eqref{eq:DPApotential} and deriving the Euler-Lagrange equation \eqref{eq:ELeq}, which leads to 
\begin{equation}\label{eq:DPAGPE}
\begin{split}
i \partial_{\tau_j} \phi_j & = - \nabla_{\mathbf{s}_j}^2 \phi_j + {\color{black}2\left( \phi_j -e^{i \theta_j}\right) } \\
i \partial_{\tau_{j'}} \phi_{j'} & = - \nabla_{\mathbf{s}_{j'}}^2 \phi_{j'} + (K-1){\color{black} \phi_{j'}} 
\end{split} \; 
\; \mbox{with} \;
\begin{cases}
z \ge 0, & (j,j')=(1,2) \\
z \le 0, & (j,j')=(2,1)
\end{cases}
\;\; (\mathbf{DPA}),
\end{equation}
{\color{black}where $\phi_j = \left| \phi_j\right| \exp(i\theta_j)$.} Note that, alternatively, these linearized equations can be obtained directly from the GP equations of \eqref{eq:GPE} by expanding the order parameters to first order in the perturbations about their bulk values, similarly to what was done in \cite{ao1998binary}. {\color{black} For studying the interface structure, it would be sufficient to limit our attention to real order parameters and real perturbations.  Nevertheless, we insist on deriving the DPA in complex function space because we have the intention to apply our results in future work to dynamical properties such as dispersion of phonon excitations and capillary wave excitations using the Bogoliubov-de Gennes formalism.} 

When the order parameters are stationary, namely $\phi_j (\mathbf{s}_j,\tau_j) = \phi_{j0} (\mathbf{s}_j)$, we obtain the DPA to the TIGPE, 
\begin{equation}\label{eq:TIDPAGPE}
\begin{split}
- \nabla_{\mathbf{s}_j}^2 \phi_{j0} + {\color{black}2\left(\phi_{j0}  -e^{i \theta_{j0}}\right) } =0  \\
- \nabla_{\mathbf{s}_{j'}}^2 \phi_{j'0} + (K-1){\color{black}\phi_{j'0} } =0
\end{split} \; 
\; \mbox{with} \;
\begin{cases}
z \ge 0, & (j,j')=(1,2) \\
z \le 0, & (j,j')=(2,1)
\end{cases}
\;\; (\mathbf{DPA}),
\end{equation}
{\color{black} where $\phi_{j0}=\left| \phi_{j0} \right|\exp(i \theta_{j0})$.}

Note that the DPA equations appear decoupled in the order parameters. However, the form of the equations depends on which bulk phase is chosen as starting point for the expansion. Since the bulk boundary conditions are different on either side of the interface, the order parameters are implicitly coupled. Of course, the apparent decoupling greatly facilitates analytical calculations. 

\subsection{DPA solutions}

For obtaining static interface profiles we can take all functions to be real, and independent of the coordinates $x$ and $y$ (translational invariance in the directions parallel to the interface) and independent of time. The solutions $\phi_j (\rho_j,\tau_j) = \phi_{j0} (\rho_j)$ that solve  the TIGPE with the suitable boundary conditions \eqref{eq:planarbc} are simple exponentials. A unique interface is obtained by matching the solutions for $z \ge 0$ to the ones for $z \le 0$ at $z=0$ with the requirement that the functions and their first derivatives be continuous at $z=0$. This leads to
{\color{black}
\begin{equation}\label{eq:stationarysolutionsDPAGPE}
\begin{split}
\phi_{j0} (\rho_j) & = 1 - \frac{\beta}{\alpha + \beta} e^{- \alpha \left|\rho_j \right|} \\
\phi_{j'0} (\rho_{j'}) & = \frac{\alpha}{\alpha + \beta}e^{- \beta \left|\rho_{j'} \right|} 
\end{split}
\; \mbox{with} \;
\begin{cases}
z \ge 0, & (j,j')=(1,2) \\
z \le 0, & (j,j')=(2,1)
\end{cases}
\;\; (\mathbf{DPA}),
\end{equation}
}
where $\alpha = \sqrt{2}$ and $\beta=\sqrt{K-1}$. Note that Ao and Chui \cite{ao1998binary} already discussed these functions in the frame-work of perturbation expansions and identified  $\xi_j/\beta$ as the penetration depth of condensate $j$ into condensate $j'$ for $j \ne j'$. 

In the following two figures we compare order parameter profiles calculated within the DPA model with numerically exact order parameter profiles solving the GPE \eqref{eq:TIGPE}. For the symmetric case $\xi_1 = \xi_2 = \xi$, Fig.1 shows the numerically exact order parameters together with their DPAs for $K=2$ (moderately weak segregation). For an asymmetric case $2\xi_1 = \xi_2$, Fig.2 shows the numerically exact order parameters together with their DPAs, also for $K=2$. Note how the DPA differs from the exact solution in featuring steeper profiles, corresponding to a smaller interface width. In spite of this quantitative difference the DPA appears to lead to a qualitatively correct interfacial structure.

\begin{figure}
\begin{center}
	\includegraphics[width=0.80\textwidth]{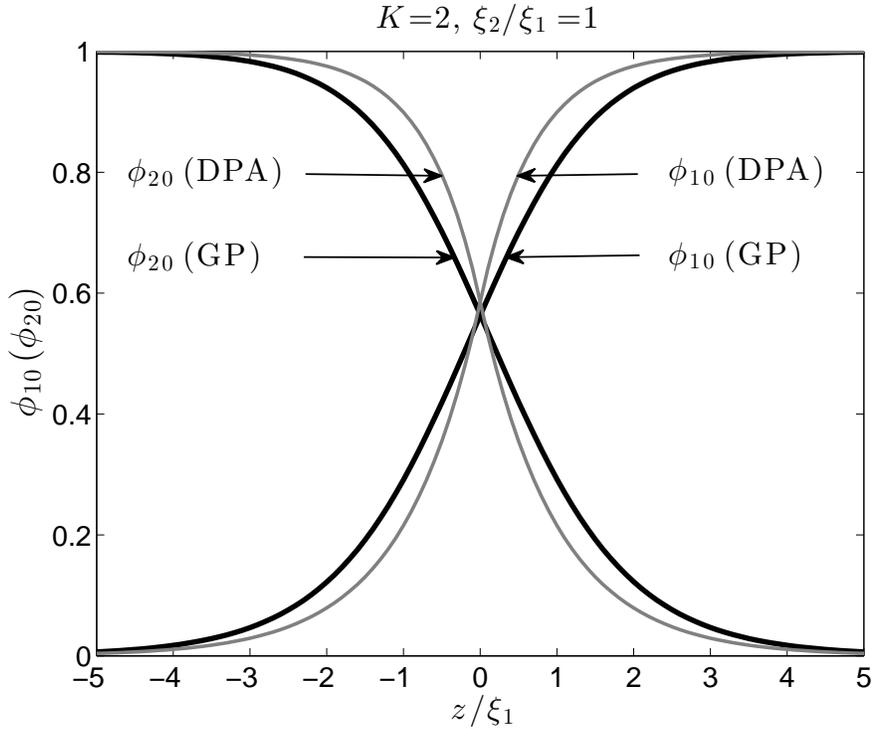}
       \caption{\label{fig1} (color online)  Interface structure for weak segregation.
Reduced order parameter profiles $\phi_{j0}$, $j=1,2$, are plotted versus $z/\xi_1$ for $K =2$ and the symmetric case $\xi_1 = \xi_2 $. The numerically exact profiles (black lines; GP) and the double-parabola approximations (grey lines; DPA) are shown. 
       }
  \end{center}
\end{figure}
\begin{figure}
\begin{center}
	\includegraphics[width=0.80\textwidth]{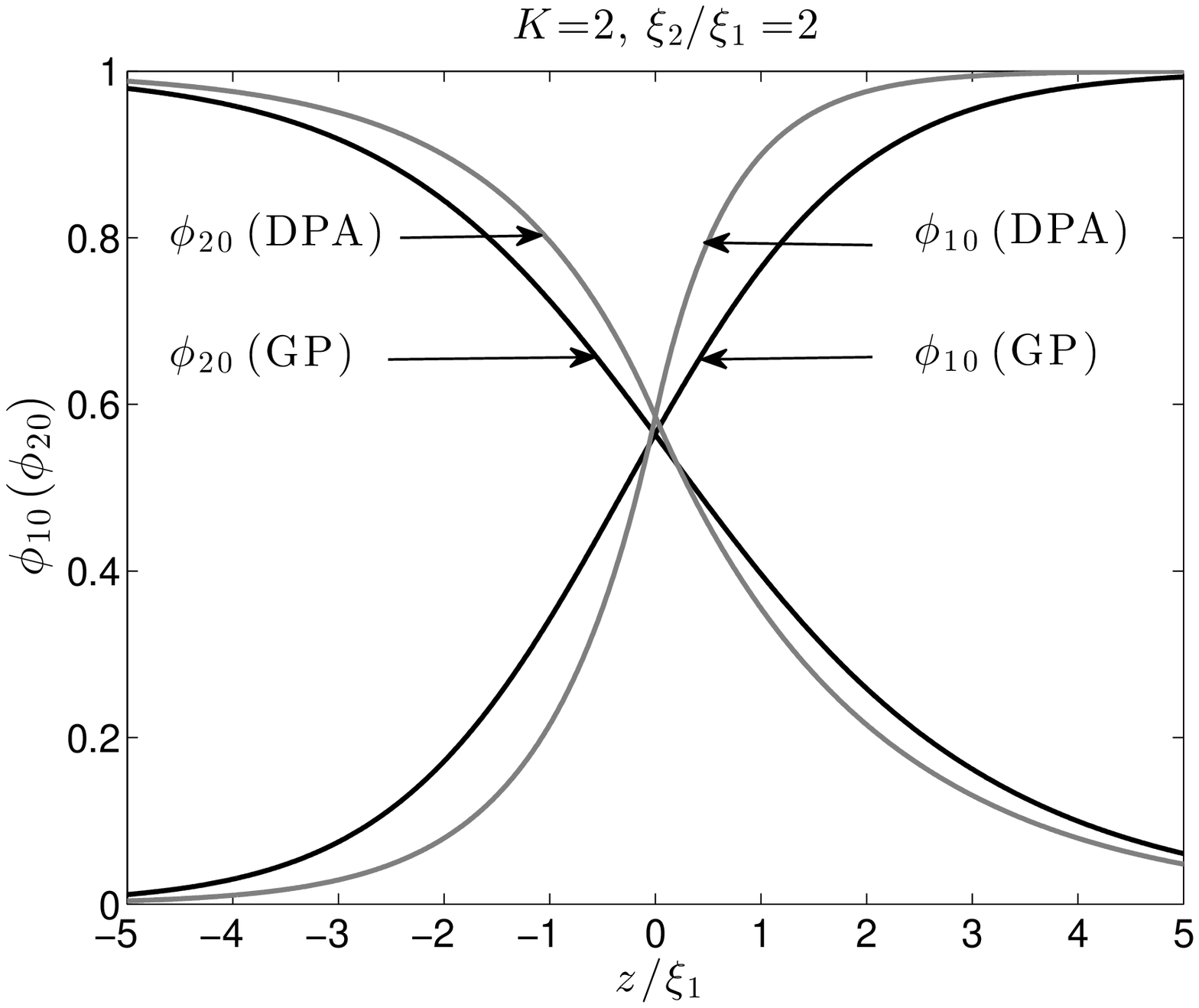}
       \caption{\label{fig2}(color online) Interface structure for weak segregation.
Reduced order parameter profiles $\phi_{j0}$, $j=1,2$, are plotted versus $z/\xi_1$ for $K =2$ and the asymmetric case $2\xi_1 = \xi_2$. The numerically exact profiles (black lines; GP) and the double-parabola approximations (grey lines; DPA) are shown. 
       }
  \end{center}
\end{figure}

\subsection{Validity conditions}

Ao and Chui \cite{ao1998binary} pointed out that, in order for these solutions \eqref{eq:stationarysolutionsDPAGPE} of the linearized GPE to be self-consistent within perturbation theory, a condition corresponding to (fairly) strong segregation must be fulfilled, which requires the penetration depth of condensate $j$ into  condensate $i$ to be smaller than the healing length of  condensate $i$, for $i=1,2$. This condition is
\begin{equation}\label{valid}
\xi_j / \sqrt{2(K-1)} <  \xi_i; \; i \neq j
\end{equation}
When this condition is not satisfied, as can typically happen for weak segregation ($K \gtrsim 1$), a term quadratic in the penetrating condensate order parameter dominates a term of first order in the deviation of the penetrated condensate order parameter from its bulk value \cite{ao1998binary}. Under such circumstances it is recommended to go beyond the DPA as far as the computation of order parameter profiles is concerned. However, for energy and interfacial tension computations we find that including this nonlinear term brings only modest improvement over the interfacial tension calculated within our DPA model, when the result is compared with the (numerically) exact result in GP theory, even in the weak segregation regime. 

Our strategy in this and forthcoming works is, and will be, to consider the DPA as a model in its own right, based on and defined by the specific potential energy density \eqref{eq:DPApotential}, and to explore its predictions. As a first example of this strategy we will, in the next section, calculate the interfacial tension within the DPA model and compare it with the interfacial tension within GP theory. As a second example we will, in  section VI, apply the DPA model for the calculation of the wetting transition in adsorbed binary BEC mixtures, and compare it with the wetting transition in GP theory. 

\subsection{Strong segregation limit}

In the limit of strong segregation $K \rightarrow \infty$, we have $\alpha/(\alpha + \beta) \rightarrow 0$ and  $\beta/(\alpha + \beta) \rightarrow 1$. The order parameters become
{\color{black}
\begin{equation}
\phi_{j0} (\rho_j) = 1 -  e^{-  \alpha \left| \rho_j \right|} ; \;\;\phi_{j'0} (\rho_{j'}) = 0,
\; \mbox{with} \;
\begin{cases}
z \ge 0, & (j,j')=(1,2)\\
z \le 0, & (j,j')=(2,1)
\end{cases}
\;\; (\mathbf{DPA})
\end{equation}
}
and we notice that there is no overlap of the condensates (and also no gap in between them; they touch at $z=0$). Complete segregation is illustrated in Fig.3 for the symmetric case $\xi_1 = \xi_2$.

\begin{figure}
\begin{center}
	\includegraphics[width=0.80\textwidth]{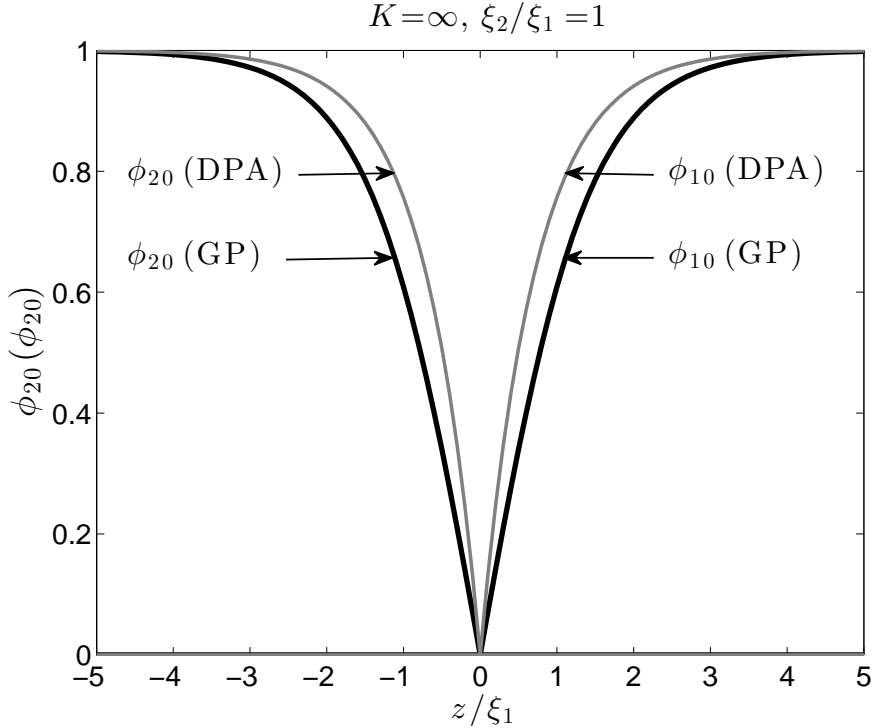}
       \caption{\label{fig3} (color online) Interface structure in the limit of strong segregation, $K \rightarrow \infty$. Reduced order parameter profiles $\phi_{j0}$, $j=1,2$, are plotted versus $z/\xi_1$ for   the symmetric case $\xi_1 = \xi_2$. The exact ``tanh" profiles (black lines; GP) and the double-parabola approximations (grey lines; DPA) are shown. 
       }
  \end{center}
\end{figure}

When we relax the strong segregation slightly so that the condensates incur a small but nonzero overlap, numerical analysis of the full GPE reveals that the interaction term $K \lvert \phi_{10} \rvert^2 \lvert \phi_{20} \rvert^2$ in the potential \eqref{eq:potential}  of the Lagrangian density \eqref{eq:dimensionlesslagrangiandensityinphi} is negligible compared to the other terms. Indeed, in spite of the fact that $K$ is large, the density overlap is so small that the product concerned is small. Now, the DPA allows us to establish analytically in which manner the interaction term vanishes in the limit $K \rightarrow \infty$. Inserting the DPA solutions for the order parameters \eqref{eq:stationarysolutionsDPAGPE} in the interaction term and calculating its value at $z=0$, we obtain
\begin{equation}\label{decoupling}
K \lvert \phi_{10} (0)\rvert^2 \lvert \phi_{20} (0)\rvert^2  =  K \left( \frac{\alpha}{\alpha + \beta} \right)^4 \propto \frac{1}{K} \rightarrow 0
\;\; (\mathbf{DPA}),
\end{equation}
which quantifies the vanishing of the interaction term for strong segregation. It is therefore clear that, in spite of the diverging coupling constant, the interaction term can be safely ignored in calculations pertaining to totally segregated condensates. Consequently, the GPE decouple in this limit. Thus, as we announced in subsection III.A, the DPA solutions can provide us with an analytical result, \eqref{decoupling}, allowing us to gain physical insight into a interesting property of the GP theory.

\section{DPA model applied to the interfacial tension}

\subsection{Grand Potential and Interfacial Tension from the GP Lagrangian}

We define a dynamical grand potential as the grand potential in equilibrium but with time-dependent order parameters, 
\begin{equation}\label{eq:grandpotential}
\Omega[\phi_1,\phi_2] =  2 P_0 \int \mathrm{d} \mathbf{r}\; \tilde{\mathcal{W}} (\phi_1,\phi_2),
\end{equation}
where  the integration is over the volume $V$ of the system and the grand potential density is defined as minus the Lagrangian density in \eqref{eq:dimensionlesslagrangiandensityinphi}, i.e.,
\begin{equation}
\tilde{\mathcal{W}} (\phi_1,\phi_2) \equiv - \tilde{\mathcal{L}} \left(\phi_1,\phi_2 \right)
\end{equation}
Consider an infinite planar interface at $z=0$ and assume translational invariance along $x$ and $y$. Without loss of generality, we assume the order parameters $\phi_{j0}$ for stationary states to be real and arrive at the grand potential for the interface,
\begin{equation}\label{eq:omega0}
\Omega_0 = \Omega[\phi_{10},\phi_{20}]
 =2 P_0 A \int  \mathrm{d} z  \left[\sum_{j=1}^{2}  \left(\partial_{\rho_j} \phi_{j0} \right)^2  + \tilde{\mathcal{V}} (\phi_{10},\phi_{20})\right],
\end{equation}
where $A$ is the interfacial area. 

To calculate $\Omega_0$, we first derive a ``constant of the motion". We multiply the TIGPE in \eqref{eq:TIGPE} by $\partial_{ z}\phi_{j0}$ and add up the two equations. We then integrate over $z$ and find
\begin{equation}\label{eq:constantmotion}
\sum_{j=1}^{2} \left[\left(\partial_{\rho_j} \phi_{j0} \right)^2 +  \phi_{j0}^2 - \frac{1}{2} \phi_{j0}^4\right]  - K \phi_{10}^2 \phi_{20}^2 =1/2,
\end{equation}
where the constant $1/2$ is obtained by considering the (bulk) boundary conditions for $\phi_{j0}$.
For order parameter profiles that satisfy the TIGPE we may substitute \eqref{eq:constantmotion} in \eqref{eq:omega0}. Then $\Omega_0$ reduces to
\begin{equation}\label{eq:Omega0reduced}
\Omega_0 = 4 P_0 A  \sum_{j=1}^{2}\int \mathrm{d} z   \left(\partial_{\rho_j} \phi_{j0} \right)^2 - P_0 V.
\end{equation}

The interfacial tension is defined as the excess grand potential per unit area,
\begin{equation}\label{eq:gamma12}
\gamma_{12} = \frac{\Omega_0-\Omega_b}{A}= 4 P_0 \sum_{j=1}^{2}\int \mathrm{d} z   \left(\partial_{\rho_j} \phi_{j0} \right)^2,
\end{equation}
which remains after the bulk grand potential $\Omega_b=-P_0 V$ of a homogeneous phase has been subtracted. Note that at bulk two-phase coexistence the bulk grand potentials are the same for each condensate. Note that expression \eqref{eq:gamma12} is valid for solutions of the TIGPE. If we wish to evaluate (nonequilibrium) interfacial tensions in GP theory for profiles that do not necessarily satisfy the TIGPE we must use \eqref{eq:omega0}.

\subsection{Interfacial tension within the DPA model}
Expression \eqref{eq:gamma12} for the equilibrium interfacial tension is independent of the form of the potential $\hat{\mathcal{V}}$. Therefore, we obtain the same expression if we start from the model defined by the DPA potential \eqref{eq:DPApotential} {\em and} use profiles that satisfy \eqref{eq:TIDPAGPE}. Note that there is no simple relation between the value of the GP interfacial tension and that of the one defined within the DPA model. They are equilibrium interfacial tensions for two different models. Note that our approach is fundamentally different from a trial-function approach in GP theory. If we wish to consider the DPA solutions as trial functions wthin GP theory, we must use \eqref{eq:omega0} and may not use \eqref{eq:gamma12}. Doing so would lead to an approximation that is far less useful than our DPA model because, for example, it would lead to an interfacial tension that diverges in the limit $K \downarrow 1$, which is physically unacceptable. 

In view of these considerations, an analytic expression for the interfacial tension within the DPA model is obtained by evaluating \eqref{eq:gamma12} in the DPA profiles \eqref{eq:stationarysolutionsDPAGPE}. We find
\begin{equation}\label{eq:gamma12DPA}
\gamma_{12}^{(\mathbf{DPA})} =  2 \sqrt{2} \frac{\sqrt{(K-1)/2}}{1+\sqrt{(K-1)/2}}P_0 (\xi_1 + \xi_2).
\end{equation}
This compact expression is insightful. It shows that, in the simplified model, the contribution from each condensate is proportional to its healing length. Furthermore, the expression interpolates between the strong and weak segregation limits by means of a function that depends only on $K$. How this DPA interfacial tension compares to the GP interfacial tension, is the question to which we now turn.  

\subsection{Comparison with exact results in GP theory}
\begin{figure}
\begin{center}
	\includegraphics[width=0.80\textwidth]{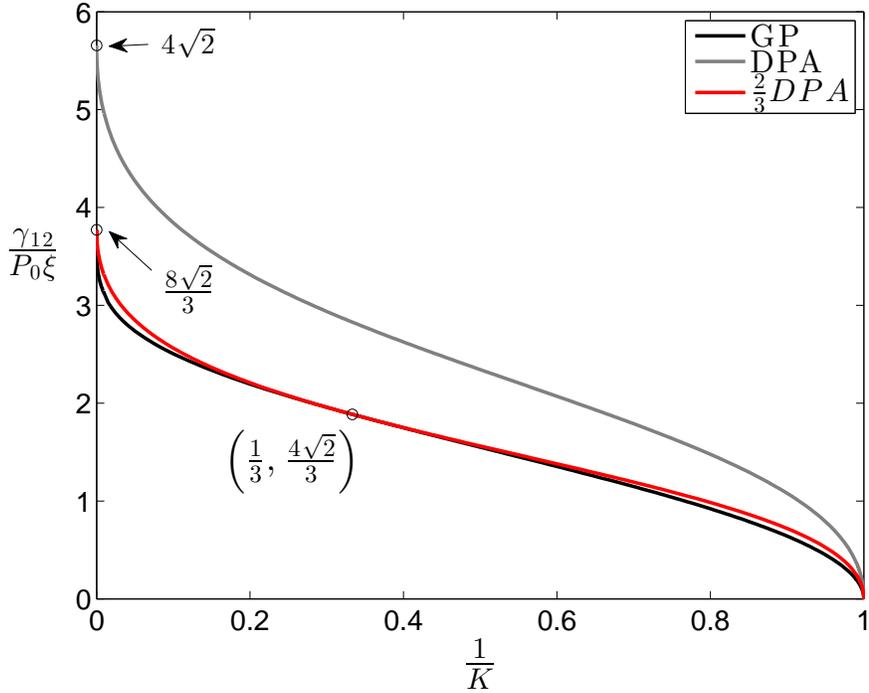}
       \caption{\label{fig4} (color online) The reduced interfacial tension $\gamma_{12}/P_0 \xi$ for the symmetric case ($\xi_1=\xi_2=\xi$) versus the inverse interaction strength $1/K$. Shown are the numerically exact solution (GP, black line), the double-parabola approximation (DPA, grey line) and the scaled DPA (red line) obtained by multiplying the DPA result by 2/3 to match the DPA and GP values at $1/K =0$. 
       }
  \end{center}
\end{figure}

%\begin{figure}
%\begin{center}
  %\epsfig{figure=phys_0b.eps,angle=0,width=180pt}
%	\includegraphics[width=0.80\textwidth]{gammaDPAcompareAoChuiIKzero.eps}
%       \caption{\label{fig4} 
%       }
%  \end{center}
%\end{figure}
\subsubsection{Strong Segregation}
In the strong segregation limit $K\rightarrow \infty$, $\gamma_{12}$  is the sum of the two wall tensions. It can be calculated directly by substituting the stationary solutions \eqref{eq:stationarysolutionsGPE} for the strong segregation limit into \eqref{eq:gamma12}, so we have
\begin{equation}\label{eq:gamma12strong}
\gamma_{12} =  \gamma_{W1} + \gamma_{W2}  = \frac{4 \sqrt{2}}{3}P_0  (\xi_1 + \xi_2),  \; \mbox{for}\; K \rightarrow \infty,
\end{equation}
where $\gamma_{Wj}$ is the wall tension for the $j$-th component \cite{Fetter}. Note that the wall tension (for a single condensate) is mathematically similar to the tension of a normal/superconducting interface in the limit of strongly type-I superconductors \cite{GL}.

Comparison with $\gamma_{12}^{(\mathbf{DPA})}$ given in  \eqref{eq:gamma12DPA} is immediate since
\begin{equation}\label{eq:gamma12strongDPA}
\gamma_{12}^{(\mathbf{DPA})} =  \gamma_{W1}^{(\mathbf{DPA})} + \gamma_{W2}^{(\mathbf{DPA})} = 2 \sqrt{2}P_0 ( \xi_1 + \xi_2), \; \mbox{for}\; K \rightarrow \infty,
\end{equation}
where $\gamma_{Wj}^{(\mathbf{DPA})} $ is the wall tension for $j$-th component. So, the ratio between the DPA and the GP result is
\begin{equation}
\frac{\gamma_{12}^{(\mathbf{DPA})}}{\gamma_{12}} = \frac{3}{2}, \; \mbox{for}\; K \rightarrow \infty.
\end{equation}

For large $K$ we can expand \eqref{eq:gamma12DPA},
\begin{equation}\label{eq:gamma12DPAexpand}
\gamma_{12}^{(\mathbf{DPA})} =  2 \sqrt{2} P_0\left(1 - \frac{\sqrt{2}}{\sqrt{K}}+ {\cal O} \left(\frac{1}{K}\right)\right)(\xi_1 + \xi_2),
\end{equation}
and observe that the interfacial tension approaches its limit with a square-root singularity (with diverging slope) in the variable $1/K$. We can compare this with the leading terms in the large-$K$ expansion of the GP interfacial tension, derived in \cite{BertVan},
\begin{equation}\label{eq:gamma12DPABertexpand}
\gamma_{12} =   P_0 \left(\frac{4 \sqrt{2} }{3} - 2.056 \frac{\sqrt{\xi_1 \xi_2}}{\xi_1 + \xi_2} \frac{1}{K^{1/4}}+ {\cal O} \left (\frac{1}{K^{1/2}}\right)\right)(\xi_1 + \xi_2).
\end{equation}
It is noteworthy that the $K^{-1/4}$ singularity as well as its amplitude 2.056... are universal in the sense that they are to some extent independent of the details of the theory. In particular, these features are common to the GP theory of BEC \cite{GP} and the earlier Ginzburg-Landau (GL) theory of superconductivity \cite{GL}. In particular, the amplitude 2.056... was first obtained by Mishonov (1.03 in his units) using GL theory \cite{Mishon}. Subsequently this was confirmed and elaborated \cite{BI} and later applied to BEC \cite{BertVan} using GP theory.

The DPA result \eqref{eq:gamma12DPAexpand} does not capture the $K^{-1/4}$ singularity and displays a $K^{-1/2}$ singularity instead. The difference in the manner the slope diverges near $1/K=0$ can be appreciated in Fig.4, in which both the DPA and the GP interfacial tension are plotted as a function of $1/K$ for the symmetric case $\xi_1 = \xi_2$. In this figure we observe that the DPA curve is similar in shape to the GP interfacial tension. If we reduce the DPA by applying an overall prefactor of 2/3, we obtain the red curve, which follows the GP curve surprisingly well. Note that the reduced DPA intersects the GP curve in one internal point, to which we will return shortly, after discussing the other interesting limit, $K \downarrow 1$.

\subsubsection{Weak Segregation}

In the weak segregation limit $K \downarrow 1$, the two condensates tend to merge. The total density of the two condensates is nearly constant and displays a small depression around the interface. In this limit $\gamma_{12}$ was calculated by Barankov \cite{barankov2002boundary}, and before him by Malomed {\em et al.} \cite{Malomed} in a different physical context, who obtained
\begin{equation}\label{eq:gamma12weak}
\gamma_{12} =  \frac{4 \sqrt{K-1}}{3} P_0 \frac{\xi_1^3-\xi_2^3}{\xi_1^2-\xi_2^2} = \frac{4 \sqrt{K-1}}{3} P_0 \frac{\xi_1^2 + \xi_1 \xi_2 + \xi_2^2}{\xi_1 + \xi_2}.
\end{equation}
When $\xi_j= \xi$, it simplifies to
\begin{equation}
\gamma_{12} = 2 \sqrt{K-1}P_0 \xi.
\end{equation}

Considering the DPA, when we take the limit $K \downarrow 1$, we find that $\gamma_{12}^{(\mathbf{DPA})}$ in \eqref{eq:gamma12DPA} approaches
\begin{equation}\label{eq:gamma12weakDPA}
\gamma_{12}^{(\mathbf{DPA})} =  2 \sqrt{K-1} P_0 (\xi_1 + \xi_2).
\end{equation}
Comparing the DPA with the GP result for the symmetric case $\xi_j =\xi$, we have
\begin{equation}
\frac{\gamma_{12}^{(\mathbf{DPA})}}{\gamma_{12}} = 2, \; \mbox{for} \; K \downarrow 1.
\end{equation}
We conclude that the DPA model describes the interfacial tension in the weak segregation regime qualitatively correctly, since it shares the correct square-root singularity at $K=1$ with the GP expression. This can also be appreciated in Figure 4.

\subsubsection{Half-segregation and the exact solution of Malomed {\em et al.}}
We already noted, when we discussed Fig. 4, that the reduced DPA for the interfacial tension intersects the GP curve in one internal point. This happens at $K=3$, for the symmetric case $\xi_1=\xi_2$. Interestingly, at this point in parameter space an exact solution to the GP equations was provided by Malomed {\em et al.} \cite{Malomed}. We have already recalled this solution in \eqref{Malomed}. Note that the order parameters are perfectly symmetric. Not only are the healing lengths equal but the healing length also equals the penetration depth since $\sqrt{2}= \sqrt{K-1}$. The profiles cross precisely half-way their bulk values, at $\phi_{j0} =1/2$. Therefore, we denote this special case by ``half-segregation". 

It is instructive to observe that the DPA solutions \eqref{eq:stationarysolutionsDPAGPE} are also perfectly symmetric in this case and display half-segregation. The interfacial tension within the DPA model is given by 
\begin{equation}\label{eq:gamma12DPAhalf}
\gamma_{12}^{(\mathbf{DPA})} =  2\sqrt{2} P_0 \xi,
\end{equation}
while the GP value is precisely two thirds of this, 
\begin{equation}\label{eq:gamma12half}
\gamma_{12} =  \frac{4\sqrt{2}}{3} P_0 \xi,
\end{equation}

\section{Application of the DPA model to wetting phenomena}

\subsection{Wetting phase boundary}
In this section we apply the DPA model to the wetting phase transition predicted for BEC mixtures adsorbed at an optical wall, using the GP theory at $T=0$ \cite{IVS,IVS2}. The wetting transition takes places when, e.g., a layer of condensate 2 intrudes between condensate 1 and the optical wall (evanescent wave emanating from a prism). The condition for a wetting transition is the following surface energy equality,
\begin{equation}\label{Antonov}
\gamma_{W1} =  \gamma_{W2} + \gamma_{12}.
\end{equation}
The curve in the $(\xi_2/\xi_1,1/K)$-plane that solves this equation is the so-called wetting phase boundary. In the hard wall limit (for a vanishing order parameter at the wall), the wetting phase boundary was established numerically \cite{IVS} and an analytical solution was reported in the second paper of \cite{BertVan}. In this limit the wetting transition is of first order (discontinuity in the first derivative of the energy). The analytic solution for the phase boundary is 
\begin{equation}\label{eq:wettinglinerelation}
\sqrt{K-1} =\frac{\sqrt{2}}{3} \left[ \frac{\xi_1}{\xi_2} - \frac{\xi_2}{\xi_1}\right].
\end{equation}
For the more physical case of a softer wall, the wetting transition was studied in \cite{IVS2}. It was found that first-order wetting as well as critical wetting are possible. 

Our aim is to derive an approximate wetting phase boundary in the hard wall limit within the DPA model and to compare it with the GP result \eqref{eq:wettinglinerelation}. To this end we first give the DPA for the wall tensions,
\begin{equation}
\gamma_{Wj}^{(\mathbf{DPA})} = 2 \sqrt{2} P_0 \xi_j,
\end{equation}
as follows from \eqref{eq:gamma12DPA} in the limit $K \rightarrow \infty$.
Inserting these and our expression \eqref{eq:gamma12DPA} for $\gamma_{12}^{(\mathbf{DPA})}$ into \eqref{Antonov}, leads to the wetting phase boundary within the DPA model,
\begin{equation}
\sqrt{ K-1} =\frac{1}{\sqrt{2} } \left[ \frac{\xi_1}{\xi_2} -1\right],\; (\mathbf{DPA}).
\end{equation}
Figure 5 shows the GP wetting phase boundary together with the DPA. Clearly, the two curves are almost coincident. Moreover, the DPA reproduces the parabolic character of the GP phase boundary near both endpoints, at $1/K=0$ and $K=1$.

\begin{figure}
\begin{center}
	\includegraphics[width=0.80\textwidth]{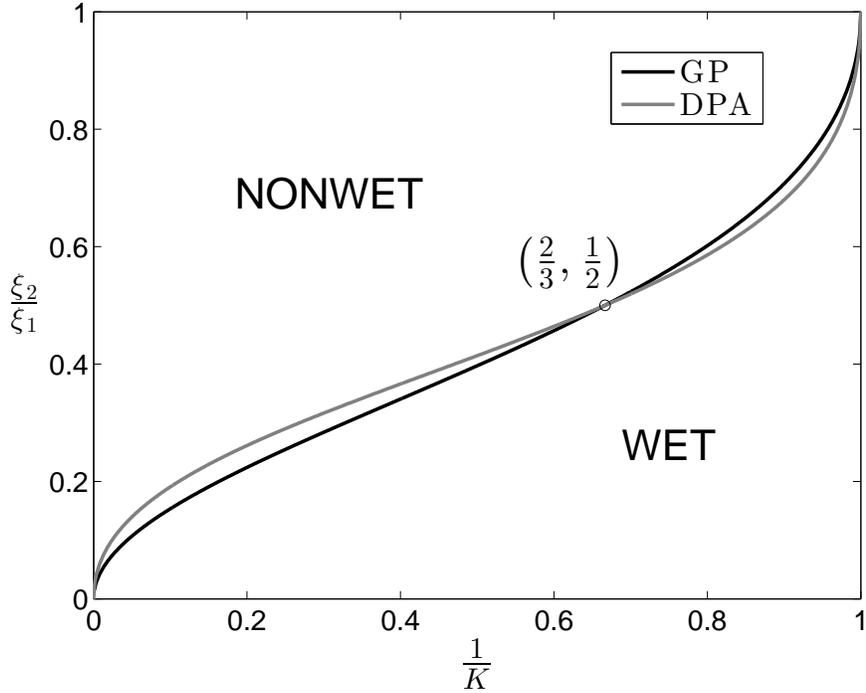}
       \caption{\label{fig5} (color online) Phase boundary for the first-order wetting transition in adsorbed BEC mixtures, in the plane of inverse interaction strength and healing length ratio. The GP solution is shown (black curve) as well as the DPA (grey curve). 
       }
  \end{center}
\end{figure}

\subsection{DPA-assisted design of an exact solution to the GP equations}
In this subsection we explain how we found the exact solution \eqref{serendipity} announced in section III.C and displayed in Fig. 6. We start by observing that the DPA intersects the GP curve precisely at $\xi_2/\xi_1 = 1/2$ and $K=3/2$ (see Fig. 5). The asymptotic behavior of the order parameters can be read off from the DPA solutions, provided conditions \eqref{valid} are satisfied. For $z>0$ we can rely on the DPA since $\xi_2/\xi_1 < \sqrt{2(K-1)}=1$. An ``up-down" symmetry occurs since $\xi_1/\sqrt{2}$ equals the penetration depth $\xi_2/\sqrt{K-1}$. For $z<0$ we cannot rely on the DPA for the approach of $\phi_{20}$ towards 1, since $\xi_1/\xi_2 >1$. In this case the approach towards the bulk density is governed by the decay length $\xi_1/(2\sqrt{K-1})$, which is longer than the length $\xi_2/\sqrt{2}$ predicted by the DPA. 

This information, together with the graphical observation that $\phi_{10}$ takes a value of about $1/\sqrt{2}$ at the point in space where $\phi_{20}$ equals 1/2, suggest that the solutions to the GPE ought to be well approximated by the following skewed profiles presented in \eqref{serendipity}, {\em which happen to solve the GPE exactly}. Note that the $z$-coordinate can be shifted so as to provide the profile crossing at $z=0$, which facilitates comparison with the DPA which intrinsically features this position as the location of the interface. The required shift is $\delta=\text{arctanh} (2-\sqrt{5})$ in units of $\sqrt{2}\, \xi$. Fig.6 shows these GP solutions together with their DPA counterparts. 
\begin{figure}
\begin{center}
	\includegraphics[width=0.80\textwidth]{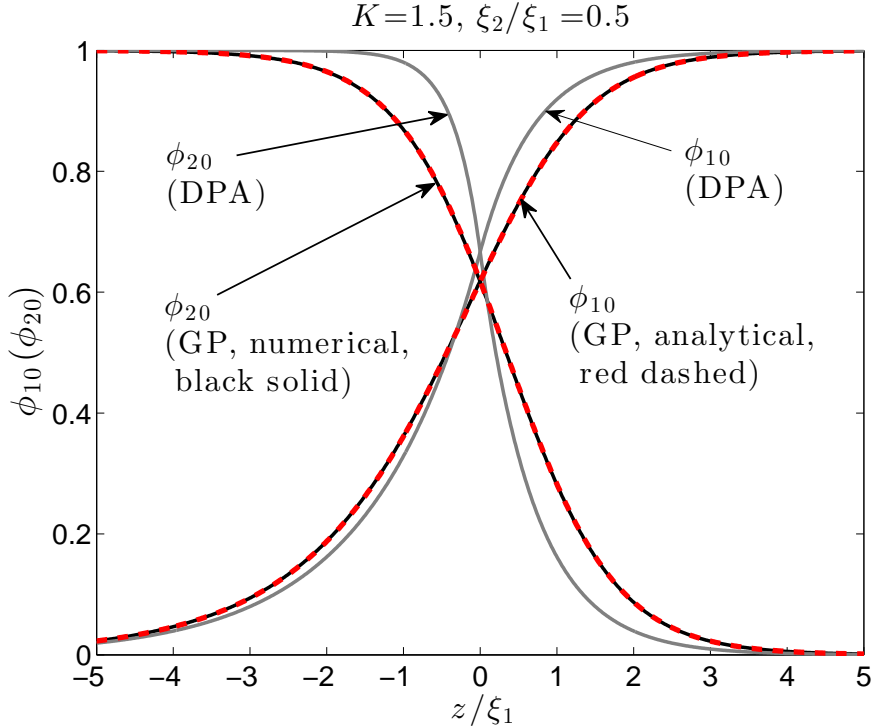}
       \caption{\label{fig6}(color online) Interface structure for an exactly solvable asymmetric case. Reduced order parameter profiles $\phi_{j0}$, $j=1,2$, are plotted versus $z/\xi_1$ for $K =3/2$ and  $\xi_1 = 2\xi_2$. The numerically exact profiles (black lines; GP), the exact solution (red dashed lines; analytical) and the double-parabola approximations (grey lines; DPA) are shown.  
       }
  \end{center}
\end{figure}

Incidentally, we note that the interfacial tension obtained for this exact solution
\begin{equation}
\gamma_{12}=   \frac{2\sqrt{2}}{3} P_0 \xi.
\end{equation}
is  2/3 of the value found within the DPA model for the same condensate parameters $K =3/2$ and $\xi_1 = 2\xi_2 \equiv \xi$.

\section{Conclusion}
In this work we accomplished two goals: i) We added an exact solution to the GP theory for interfaces in BEC binary mixtures. To our knowledge this is the first exact solution for an asymmetric system (with unequal healing lengths) and at finite inter-species repulsion strength $K$. We have been able to find this solution guided by information gathered by solving a simpler but related model, the so-called DPA. ii) We defined and developed the ``DPA model". We first derived the DPA for the potential energy density in the Lagrangian by expanding the order parameters about their bulk values and keeping the deviations into account to second order. Locating the interface center at $z=0$, we next derived the DPA for the GP equations in each half-space, $z<0$ and $z>0$. The solutions and their first derivatives were then matched at $z=0$. This led to unique simple analytical  solutions that can be used efficiently to uncover and understand properties of GP theory. 

The power of the DPA model lies in its capacity to provide systematically analytical expressions for many physical quantities of two segregated BECs. We know of no other methods capable of doing this. Here, the excess surface energies at a hard wall and at the interface were evaluated within the DPA model. This provided a compact and useful expression for the interfacial tension. As an application we derived the wetting phase boundary within DPA and obtained good agreement with the GP solution. Moreover, the DPA provided crucial hints facilitating  a successful guess of an exact solution to the GP equations for an asymmetric case $\xi_1 \neq \xi_2$. Clearly, the DPA model is a practical and broadly applicable tool for exploring the physics of a more complicated model. In the future we plan to use the DPA frame-work to derive, and to get physical insight in, an approximate dispersion relation for capillary waves on the interface, which takes into account the finite thickness and the structure of the interface.

%% If you have acknowledgments, this puts in the proper section head.
\begin{acknowledgments}
%% put your acknowledgments here.
N.V.T and T.H.P are supported by the Vietnam National Foundation for Science and Technology Development (NAFOSTED) and J.O.I. and C.-Y.L. by FWO Flanders under Grant Nr. FWO.103.2013.09 within the
framework of the FWO-NAFOSTED cooperation. J.O.I. and C.-Y.L. are furthermore supported by  KU Leuven Grant OT/11/063. The authors thank Hans Hooyberghs, Mehran Kardar and Todor Mishonov for discussions.
\end{acknowledgments}


\end{document}